\input psfig.sty
\documentstyle{neu}
\begin{document}

\title{%
OPTICAL OBSERVATIONS OF ISOLATED NEUTRON STARS.}

\author{Roberto P. MIGNANI \\
{\it   Max Planck Institute f\"ur Extraterrestrische Physik, Postfach 1603 Giessenbachstrasse, D85740
Garching, Germany, rmignani@xray.mpe.mpg.de}\\
}

\maketitle



\section{Introduction.}

\noindent
After 30 yrs, the discovery of an Isolated Neutron Stars (INS) at optical wavelengths is no longer a surprise.
Up to now, nine INSs have been associated to an optical 
counterpart (see Table 1 for an observational overview). 
Young INSs
are relatively bright and identified through the detection of optical pulsations.
For older  INSs, the much fainter optical luminosity 
prevents in most cases the source timing and the identification just relies on
the positional coincidence with a field object as well as on its  peculiar colors.
However, for some of the closer $(d \le 500 ~ pc)$ INSs other, independent, 
pieces of evidence can be obtained, e.g., by comparing the angular displacement of the optical 
counterpart, if any, with the proper motion of the radio pulsar (Mignani, Caraveo \& Bignami,
1997a - MCB97a).
Our understanding of the optical properties of INSs is
limited by their intrinsic faintness.  Indeed, only for the Crab, accurate,  
medium-resolution, optical spectroscopy is available
(Nasuti et  al, 1996). For few more cases (PSR0540-69, Vela, PSR1509-58, PSR0656+14
and Geminga), which are too faint to be within the spectroscopic  capabilities of the
present generation  of ground/space-based instruments, the spectral information just
relies on multicolour photometry (Nasuti et al, 1997; Mignani et al, 1998;  Pavlov et
al, 1997; Bignami et al, 1996).  For the rest of the database
(PSR0950+08, PSR1929+10  and PSR1055-52), only one-or two-band detections are available 
(Pavlov et al 1996; MCB97b).

\begin{table}[t]
\caption{Summary of the existing database for all the optically identified INSs, sorted according to their
spin-down age $\tau$.  Column 3 lists the identification evidence obtained either 
from optical pulsations (P) or proper motion of the optical counterpart (PM). 
The numbers correspond to the accuracy of the positional coincide in arcsec. 
Successful or unsuccessful (-) timing and spectroscopy, both from the ground (G) and 
from the HST (S), is listed in
Columns 4,5. Multicolor photometry (in italics for HST) is given in columns 6-10.}

\begin{center}
\begin{tabular}{l|l|l|llllllll} \hline
{\em Name} & $ Log \tau$ &  {\em ID} & {\em Tim} & {\em Spec}  & {\em I} & {\em R} & {\em V} &  {\em B} & {\em U}
& \\ \hline
Crab    & 3.1 & P $^{1}$    & G,S & G    & 15.63 & 16.21 & 16.65 & 17.16 & 16.69   \\ 
0540-69 & 3.2 & P $^{2}$    & G,S & G,S  & 21.5  & 21.8  & {\it 22.5}  & 22.7  & 22.05    \\
1509-58 & 3.2 & 0.35$^{3}$  & G(-)& G(-) & 19.8  & 20.8  & 22.1  & 23.8	&            \\ 
Vela    & 4.1 & P $^{4}$    & G   &      &       & 23.9  & {\it 23.6}  & 23.9  & 23.8      \\ \hline 
0656+14 & 5.0 & P$^{5}$     & G   &      & 23.8  & 24.5  & {\it 25}  & {\it 24.8}    & {\it 24.1}        \\
GEMINGA & 5.5 & PM$^{6}$    & G   &	   & $\ge 26.4$ & 25.5  & {\it 25.5}  & {\it 25.7}   & {\it 24.9}
    \\
1055-52 & 5.7 & 0.1$^{7}$   &     &      &       &   &  &   & {\it 24.9}     \\ \hline  
1929+10 & 6.5 & 0.39$^{8}$  &     &	   &       &   &	 &$\ge$ {\it 26.2} & {\it 25.7}     \\
0950+08 & 7.2 & 1.83$^{8}$  &     &      &       &	 &  &    & {\it 27.1}	&   \\ \hline
\end{tabular}
\end{center}

$^{1}$Cocke et al, 1969; 
$^{2}$Shearer et al, 1994; 
$^{3}$Caraveo et al, 1994; 
$^{4}$Wallace et al, 1977;
$^{5}$Shearer et al, 1997; 
$^{6}$Bignami et al 1993; 
$^{7}$MCB97b; 
$^{8}$Pavlov et al, 1996
\end{table}
\section{Young INSs.}

\noindent
Young ($\tau \sim 10^{3-4}~yrs$) INSs, hereafter YINSs, 
are relatively bright with an optical luminosity $L_{opt} \sim 10^{33} erg/s$ 
i.e. $\sim 10^{-5}$ of their spin-down power $\dot E$. 
Apart from PSR1509-58 (Mignani et al, 1998), they are all detected as 
optical pulsars. The optical emission of YINSs is powered by energetic electron interactions in the
magnetosphere and is characterized by flat, synchrotron-like spectra (Nasuti et
al, 1997). This must be the case
also for PSR1509-58, although any precise modelling is hampered by the heavy 
interstellar absorption (Mignani et al, 1998).

\section{Middle-Aged INSs.}

\noindent
Middle-Aged INSs ($\tau \sim 10^{5}~yrs$) - MINSs-, 
are much fainter ($L_{opt} \sim 10^{27-28} erg/s$) and optical pulsations 
have been observed only recently for PSR0656+14 and Geminga (Sherear et al,
1997a,b). 
Describing the optical luminosity of MINSs in terms of a single model
is not straightforward since different emission processes may become important (MCB97c).  
For example, the optical magnetospheric emission might have faded enough to
make it detectable the thermal emission from the cooling neutron star surface at
a temperature in the $10^{5}- 10^{6} ~ K$ range (see e.g. Nomoto \&
Tsuruta, 1987), in agreement with X-ray observations (e.g. Becker \& 
Tr\"umper, 1997).  
Indeed, recent multicolor photometry of Geminga (MCB98) 
clearly suggests thermal radiation from the neutron star surface as the most likely
origin of its underlying optical emission. At the same time, a wide emission
feature @ $\sim 6,000 \AA$, which can be attributed to cyclotron emission from light ions in 
the neutron star's atmosphere, appears superimposed on the Rayleigh-Jeans 
continuum of the soft X-ray planckian.
However, for PSR0656+14, indeed the youngest of the three, multicolor photometry 
(Pavlov et al, 1997) show a different spectral behaviour with the
optical flux consistent with a steep power law plus
an additional hot blackbody component.  
For PSR1055-52, the HST/FOC point (MCB97b) falls close to the RJ extrapolation of 
the ROSAT thermal spectrum, but certainly more data are required to investigate the 
optical emission processes. 

\section{Old INSs.}

\noindent
Very little is known on the optical emission of the even fainter ($L_{opt} \sim 10^{26} erg/s$) 
Old ($\tau \ge 10^{6}~yrs$) INSs (OINs). For both PSR1929+10 and PSR0950+08, the measured magnitudes (Pavlov et al, 1996) 
are clearly unconsistent with the optical extrapolation of the polar caps/magnetospheric soft X-ray spectra.
Again, the optical flux could be due to thermal emission from a cooler neutron star surface.

\section{Conclusions.}

\noindent
As inferred from Fig.1, a decay of the optical luminosity of INSs is evident. Still to be 
established are the effective rate of such a decay and whether it is followed by a turnover in the optical
emission mechanisms.
For YINSs, the optical output is certainly magnetospheric. However, as originally proposed by Pacini (1971),
optical magnetospheric emission is expected to fade away rapidly. 
Indeed, the slightly older Vela is underluminous ($L_{opt} \sim 10^{28} erg/s$), 
with only $\sim 10^{-9}$ of its spin down power emitted in the optical.  
This fading could thus give way to the fainter thermal emission from the cooling neutron star's 
surface. Indeed, the optical emission of MINSs appears to be either purely thermal or 
the combination of thermal/non-thermal components. Thermal emission could finally take 
over the non-thermal one in the case of the OINSs. 
Of course, any evolutionary picture is far from being settled. More 
observations, to be performed with the next generation of ground/space based telescopes, 
are required both to obtain new identifications and to gather additional spectral information.

\begin{figure}[h]  
\centerline{\hbox{ 
\psfig{figure=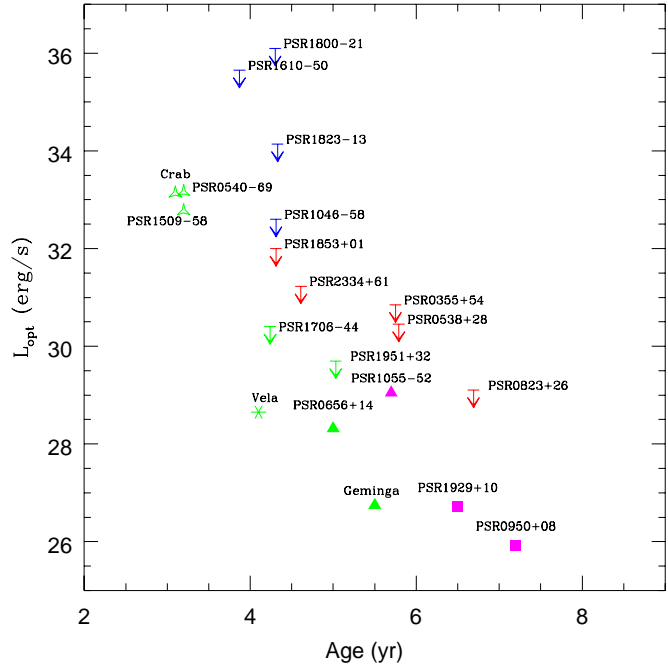,height=9cm,clip=}
}}
\caption{Optical luminosities of INSs vs their spin-down age. Apart from Geminga, 
the values have been computed for the nominal radio distances.
YINSs (open triangles), MINS (full triangles) and OINSs (filled squares) are marked.
Vela is marked by an asterisk.  Present upper limits are also included.}
\label{}
\end{figure}

%
%
%
%

\vspace{1pc}
\re Becker, W. \& Tr\"umper, J. 1997, {\it A\&A 326, 682}

\re Bignami, G.F., Caraveo, P.A. and Mereghetti, S., 1993 {\it Nature 361, 704}

\re Bignami, G.F., Caraveo, P.A., Mignani, R., Edelstein, J. and Bowyer, S. 1996, 
{\it Ap.J. Lett. 456, L111} 

\re Caraveo, P.A., Mereghetti, S. and Bignami, G.F., 1994 {\it Ap.J. 423, L125}

\re Cocke W.J., Disney M.J. and Taylor D.J., 1969, {\it Nature 221, 525}
 
\re Mignani, R., Caraveo, P.A. and Bignami, G.F. 1997a, {\it The Messenger 87, 43} 
 
\re Mignani, R., Caraveo, P.A. and Bignami, G.F. 1997b,  {\it Ap.J. 474,L51} 

\re Mignani, R., Caraveo, P.A. and Bignami, G.F. 1997c, {\it Adv. in Sp. Res. - in
press}

\re Mignani, R., Caraveo, P.A. and Bignami, G.F. 1998, {\it submitted}

\re Mignani, R. et al 1998, - in preparation

\re Nasuti, F. P., Mignani, R., Caraveo, P. A. and Bignami, 
G.F. 1996, {\it A\&A 1996, 314,849} 

\re Nasuti, F.P., Mignani, R., Caraveo, P.A. and Bignami, 
G.F. 1996, {\it A\&A 1997, 323,839} 

\re Nomoto, K.\& Tsuruta, S., 1987 {\it Ap.J. 312, 711}

\re Pacini, F., 1971 {\it Ap.J. 163, L17}
 
\re Pavlov, G.G., Stringfellow, G.S. and Cordova, F.A., 1996 {\it Ap.J.   467,370} 

\re Pavlov, G.G., Welty, A.D. and Cordova, F.A., 1997 {\it Ap.J. 489, L75}

\re Shearer, A. et al, 1994 {\it Ap.J. 423, L51}

\re Shearer, A. et al, 1997a {\it Ap.J. 487, L181}

\re Shearer, A. et al, 1997b {\it Ap.J. IAUC 6787}

\re Wallace, P.T. et al, 1977 {\it Nature 266, 692}

\vspace{1pc}

\newpage
%
%

\chapter*{ Entry Form for the Proceedings }

\section{Title of the Paper}

{\Large\bf %

Optical observations of Isolated Neutron Stars. 

}

\section{Author(s)}

\newcounter{author}
\begin{list}%
{Author No. \arabic{author}}{\usecounter{author}}

\item %
\begin{itemize}
\item Full Name:                Roberto Mignani
\item First Name:               Roberto 
\item Middle Name:              Paolo  
\item Surname:                  Mignani 
\item Initialized Name:         R. P. Mignani 
\item Affiliation:              Max Planck Institute f\"ur Extraterrestrische Physik, Garching 
\item E-Mail:                   rmignani@xray.mpe.mpg.de 
\item Ship the Proceedings to:  MPE, Postfach 1603 Giessenbachstrasse, D85740 Garching, Germany 
\end{itemize}


\end{list}

\end{document}